\documentclass[aps,prl,twocolumn,amssymb,showpacs]{revtex4}
\usepackage{graphicx}
\usepackage{amsmath}
\usepackage{txfonts}

\newlength{\figwidth}
 \divide\figwidth by 2
\begin{document}

\title{Magnetic order in Tb$_2$Sn$_2$O$_7$ under high pressure:
from ordered spin ice to spin liquid and antiferromagnetic order.}
\author{I. Mirebeau$^1$, I. Goncharenko$^1$, H. Cao$^1$, A. Forget$^2$.
\\
}
\affiliation{
$^1$Laboratoire L\'eon Brillouin, CEA-CNRS, CE-Saclay, 9191
Gif-sur-Yvette, France.}
\affiliation {$^2$Service de Physique de l'Etat Condens\'e,
CEA-CNRS, CE-Saclay,  91191 Gif-Sur-Yvette, France.}

\begin{abstract}
We have studied the Tb$_2$Sn$_2$O$_7$ frustrated magnet by
 neutron diffraction under isotropic pressure  of 4.6 GPa, combined with uniaxial pressure of 0.3
 GPa, in the temperature range 0.06~K$<$T$<$100~K. Magnetic order persists under pressure but
 the ordered spin ice structure stabilized at ambient pressure below 1.3~K partly
 transforms into an antiferromagnetic one.  The long range ordered moment at 0.06~K is reduced under pressure, which is interpreted
by a pressure induced enhancement of the spin liquid fluctuations.
 Above the ordering transition, short range spin correlations
 are affected by pressure, and ferromagnetic correlations are suppressed.
The influence of pressure on the ground state is discussed considering
both isotropic and stress effects.
\end{abstract}

 \maketitle

Original orders can be found in geometrically frustrated
magnets, with magnetic ground states showing similarities
with the liquid, ice, and glassy states of matter. Among
them, the spin ices attract increasing
attention. Their magnetic ground state can be mapped to that of real
ice\cite{Bramwell01,Bramwell012} and possesses the same entropy\cite{Ramirez99}. 
Whereas in many magnets antiferromagnetic (AF) first neighbor interactions are geometrically frustrated, 
in spin ices this occurs for ferromagnetic (F) interactions, when combined
with the local Ising-like anisotropy of the magnetic moments.
In  pyrochlore spin ices R$_2$Ti$_2$O$_7$, where the rare earth ion R is Dy or Ho,
 the  effective ferromagnetic interaction between R moments results from
 AF superexchange and F dipolar interactions.
 The strong crystal field anisotropy
 of the 
  Dy$^{3+}$ or Ho$^{3+}$ moments constrains the moments to lie along the $<$111$>$ local axes
 connecting the center of each tetrahedron to its corners.

The Terbium pyrochlores show a more complex but even richer
behavior, due to the smaller anisotropy of the Tb$^{3+}$ ion \cite{Gingras00,Mirebeau07}, and to
the fact that superexchange and dipolar interactions between
near-neighbor Tb$^{3+}$ ions nearly compensate. Tb$_2$Sn$_2$O$_7$
(TSO) is an intriguing example of an ordered spin
ice\cite{Mirebeau05}.
Contrary to classical spin ices which do not order at large
scale, here the four tetrahedra of the unit cell have identical moment orientations,
yielding long range order (LRO) 
below T$_{\rm I}$=1.3(1) K. An strong increase of the
correlation length and ordered magnetic moment occurs at T$_{\rm C}$=0.87(2)K, together with a peak in
the specific heat.
The nature of this LRO, and its coexistence with spin fluctuations in the ground state is highly debated \cite{Bert06,Dalmas06,Rule07,Chapuis07,Mirebeau08,Giblin08}.


Whereas TSO shows an ordered ground state, the sibling compound Tb$_2$Ti$_2$O$_7$ (TTO)
does not order at ambient pressure, showing liquid like fluctuations down to 50 mK \cite{Gardner99}. Recent
theories\cite{Molavian07} suggest that
quantum fluctuations are responsible for its ground
state. These different ground states arise from the lattice expansion induced by Ti/Sn substitution.
Taking compressibility\cite{Kumar06,Apetrei07} into account,
 substitution of Ti for Sn corresponds to a
negative chemical pressure of 12-15 GPa.

 In TTO, the spin liquid (SL) ground state is
 strongly suppressed under pressure and
 long range AF order is induced
\cite{Mirebeau02}. Further single crystal measurements showed that this effect mainly arises from the lattice distortion induced by a uniaxial stress, allowing one to tune the ordered magnetic moment and N\'eel temperature by the stress orientation\cite{Mirebeau04}.

 Here we report the first measurements of TSO under pressure. We studied the pressure induced state in a powder sample, both in the paramagnetic region up to 100~K and in the ordered spin ice region down to 0.06~K. Considering the pressure behavior of TTO, two scenarios could be $\it{a~priori}$ predicted for the effect of pressure in TSO: i) a $"$melting$"$ of the spin ice long range order (LRO), due to the decrease of the lattice constant, yielding a spin liquid state similar to that in TTO at ambient
pressure ii) a change from the spin ice LRO with
F ordered moment to another LRO structure of
 AF character, as in TTO under stress. These two behaviors should be connected with the nature (isotropic
 or uniaxial) of the applied pressure.

By
 combining a high isostatic pressure of 4.6 GPa
 with a uniaxial stress of 0.3(1)~GPa, 
 we find that the ordered spin ice structure partly transforms into an AF
 one. Both orders
 coexist at 0.06~K but their transition temperatures slightly differ.
 The LRO moment at 0.06~K is reduced with respect
 to its ambient pressure value,
suggesting that the spin liquid ground state is favored by pressure.
 In the paramagnetic region, short range order (SRO)
 between Tb moments is also affected by pressure.
 The ferromagnetic correlations 
are suppressed, whereas AF first neighbor correlations remain
unchanged.
 The pressure induced ground state is discussed considering the influence of an isostatic compression
  and stress-induced lattice distortion separately.

\begin{figure} [h]
\includegraphics* [width=\columnwidth] {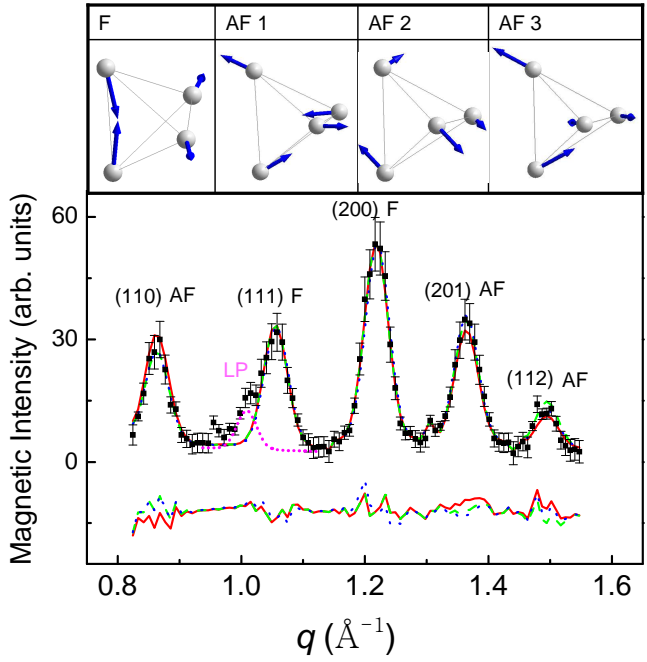}
\caption{(Color on line) Magnetic diffraction pattern in Tb$_2$Sn$_2$O$_7$ at
0.06~K under a pressure of 4.6 GPa (P$_{u}$$\sim$0.4 GPa).
 A spectrum at 4.5~K was subtracted. Lines are refinements involving both F and AF
 structures. Solid red (model 1), dashed green (model 2) and dotted blue (model 3).
 Top: spin structures under pressure in one tetrahedron. From left to right \textbf{k}=0 F structure,
\textbf{k}=(0,0,1) AF structures (models 1, 2, 3).}
\label{fig1.eps}
\end{figure}
 A  Tb$_2$Sn$_2$O$_7$ powder sample
  was inserted in a sapphire anvil
 cell, with an isostatic pressure component P$_{i}$=4.6(1)~GPa.
 High pressure neutron diffraction
 patterns
 were recorded at the diffractometer G6-1 of the Laboratoire L\'eon Brillouin\cite{GoncharenkoG61},
 with an incident neutron wavelength $\lambda$=4.74 \AA.
 Two experimental set-ups were used.
 In one set-up, the sample was mixed with a pressure transmitting medium (40 volumic\% NaCl),
 yielding a uniaxial component P$_{u}$$\sim$ 0.2 GPa along the axis of the
 pressure cell. The cell was inserted in a helium cryostat and diffraction patterns
were recorded at temperatures between 100~K and
  1.5~K to measure the SRO. Diffraction patterns were also recorded at ambient
  pressure on G6-1 and G4-1 ($\lambda$=2.426 \AA) spectrometers  
  for comparison.
  In the other set-up, no transmitting medium was used, to maximize the sample volume and
   increase P$_{u}$ to $\sim$ 0.4~GPa. The
   cell was fixed on the dilution insert of a cryostat, and diffraction patterns
    recorded between 0.06~K and 4.5~K to measure the LRO.
   Specific care was taken to
reduce the background. 
    Magnetic patterns measured in the momentum
    transfer range 0.8$<$q$<$1.6 \AA$^{-1}$ were obtained by
    subtracting a pattern measured at
    100~K and 4.5~K for SRO and LRO respectively. The
    ordered Tb moments were calibrated by
    measuring the intensity of the (222) nuclear peak.

 The magnetic pattern of Tb$_2$Sn$_2$O$_7$ under pressure at 0.06~K (Fig.\ref{fig1.eps})
 shows the coexistence of two families of Bragg peaks.
 Those of the face centered cubic lattice, indexed in the
 cubic unit cell of {\itshape Fd$\overline{3}$m} symmetry with a propagation vector
\textbf{k}=0, correspond to a magnetic order akin to that in TSO
at ambient pressure. Those of the simple
cubic lattice which appear under pressure show an AF
order indexed in the cubic unit cell by a propagation
vector \textbf{k}=(0,0,1). A similar AF structure was observed in
the pressure induced state of TTO. A small extra peak is observed
  for q=1.01~\AA$^{-1}$, nearby the (111) peak. It is attributed to a long period (LP) structure
with a larger unit cell than the cubic one, also observed in
TTO under pressure.The magnetic LRO structures which coexist under pressure were analyzed separately, since they do not yield Bragg
 reflections at the same positions. The intensities of the q=1.01\AA$^{-1}$ and (111)
 peaks which partly overlap were determined by fitting these peaks by
Gaussian shapes.

\begin{figure} [h]
\includegraphics* [width=0.8\columnwidth] {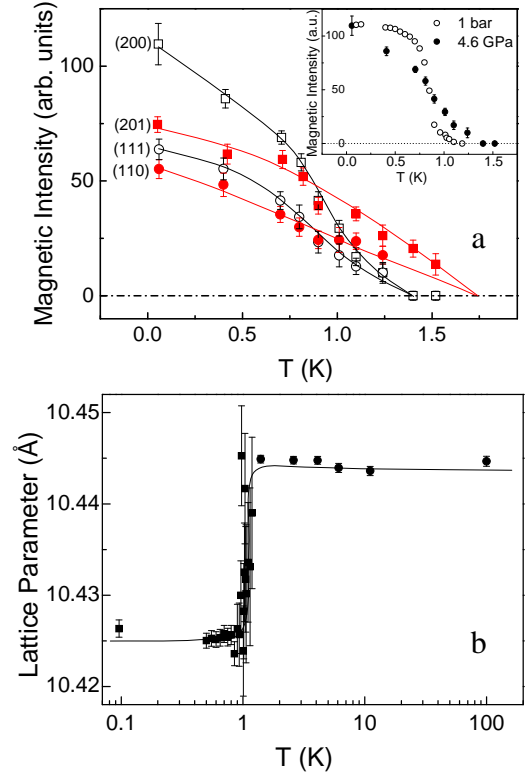}
\caption{(Color on line) (a) integrated intensity of the F ($\square$,$\circ$ 
) and AF ($\blacksquare$,$\bullet$) peaks 
versus temperature under pressure
(P$_{i}$=4.6~GPa, P$_{u}$=0.4~GPa).
Lines are guides for the eye. In inset: (200) peak
intensities at ambient and under pressure, scaled at 0.06~K. 
(b) temperature dependence of the lattice constant
at ambient pressure.} \label{fig2.eps}
\end{figure}

  The \textbf{k}=0 structure, quoted F for simplicity (Fig.~\ref{fig1.eps}, top left), has 4 identical tetrahedra in the cubic
 cell. The local spin structure in a tetrahedron was
 described
 by the same irreducible representation of the space group I4$_1$$/$amd
 as at ambient pressure\cite{Mirebeau05}.
 To refine this structure, the only parameters are
 the ordered Tb moment M$_F$ (the same for all Tb ions, determined by absolute calibration) and its canting angle
 $\alpha$ with respect to the $<$111$>$ local
 anisotropy axis. The refinement at 0.06~K (R= 1$\%$) yields M$_F$=3.3(3)\,$\mu_B$ and
  $\alpha=28(1)^\circ$ to be compared with the ambient pressure values M$_F$=5.9(1)\,$\mu_B$ and
  $\alpha=13^\circ$. So, under pressure the Tb moments in the \textbf{k}=0 structure decrease and turn away from their local easy axis. The magnetization evaluated to
  37\% of M$_F$ or 2.2\,$\mu_B$$/$Tb
  at 1~bar\cite{Mirebeau05} is reduced to 0.4(1)\,$\mu_B$ under pressure.

 The analysis of the \textbf{k}=(0,0,1) structure is more
 intricate. Here, in the cubic unit cell, two tetrahedra have identical
 orientations of the magnetic moments,
  and two tetrahedra have reversed orientations. This structure has no ferromagnetic component.
  To determine the possible local spin structures inside a tetrahedron (Fig.~\ref{fig1.eps} top right), we first searched
  for structures described by an irreducible
  representation of the I4$_1$$/$amd space group, as for TSO at ambient pressure. A possible structure (R=12\%) is given by
  model 1, which involves two couples of Tb moments of 2.6(1)\,$\mu_B$ with
  canting angles of 27(1)$^\circ$ and 33(1)$^\circ$. 
  Slightly better models were found by relaxing the symmetry constraints
  and/or the relative values of the moments in a tetrahedron. Model 2 (R=6\%) is derived from
  that
  proposed for TTO under pressure, with moments directions along $<$110$>$ axes,
  and one AF pair of Tb
  moments, which could be favored by a stress along a $[$110$]$ axis 
  \cite{Mirebeau04}.
  Model 3 (R=3\%)
  assumes that all moments have the same value (2.8(1) $\mu_B$), 2 moments have canting angles of 28(1)$^\circ$,
  and the other two make an AF pair.
   Actually, the small number of magnetic peaks and the numerous
   possible
   models 
    do not
   allow us to draw a definite conclusion. Importantly,
   whatever the solution considered, the average Tb
   moment keeps at 2.7(2)\,$\mu_B$
   at 0.06~K. 
\begin{figure} [h]
\includegraphics* [width=0.8\columnwidth] {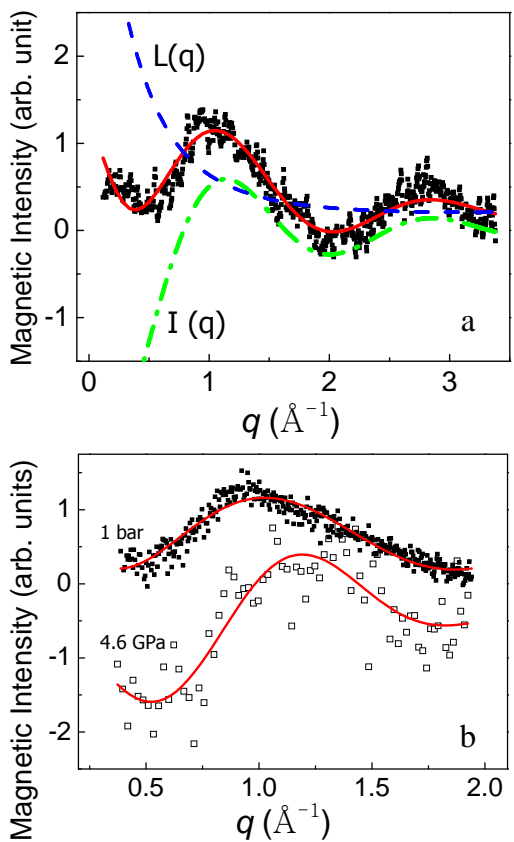}
\caption{(Colour on line) Magnetic SRO patterns at 1.5~K; a
pattern at 100~K was subtracted.
(a) ambient pressure. The solid line is a fit as described in text. Dash-dotted (dashed)
line shows the liquid like (Lorenzian) component. (b)
 SRO pattern measured at ambient pressure and under pressure
 (P$_{i}$=4.6~GPa, P$_{u}$=0.2~GPa) in the same q-range. Solid lines are fits as described in text. } \label{fig3.eps}
\end{figure}

By measuring the temperature dependence of the magnetic Bragg peaks
(Fig. \ref{fig2.eps}a), we can determine the transition temperature. As for the F structure, the value T$_{\rm C}$= 1.2(1)~K is
close to the upper transition in TSO at ambient pressure which situates at
1.3(1)~K. One notices that at ambient pressure, the order parameter
(Inset Fig. \ref{fig2.eps}) shows a steep variation at the lower transition of 0.87~K.
This suggests some first order character of the transition, supported by the
small anomaly of the lattice constant (Fig. \ref{fig2.eps}b). Under pressure, the T dependence of
the magnetic intensity is strongly smeared, without any anomaly.
  As for  the AF structure, it collapses at a slightly higher temperature T$_{\rm AF}$=1.6(1)~K.
\begin{figure} [h]
\includegraphics* [width=0.8\columnwidth] {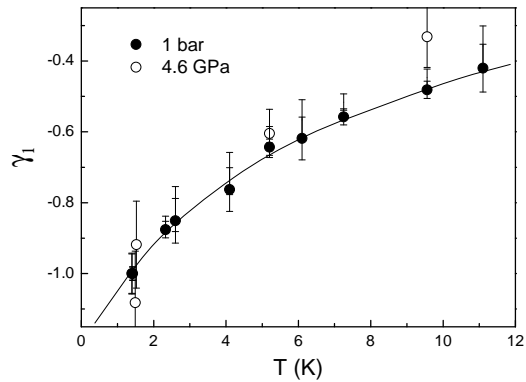}
\caption{Temperature dependence of the first neighbor correlations
in the paramagnetic region. Solid line is a guide to the eye. 
}
\label{fig4.eps}
\end{figure}

 We now discuss the influence of pressure on the SRO, as measured just above the transition.
 Difference patterns at 1.5~K with respect to T= 100~K  clearly show magnetic SRO, as broad modulations of the intensity versus the momentum transfer (Fig.3a).
 The SRO intensity I$_{SRO}$ changes under pressure, as shown in Fig. \ref{fig3.eps}b.
 This change was
 analyzed in a semi-quantitative way.
  At ambient pressure, I$_{SRO}$ was fitted by the expression I$_{SRO}$ =F$^2$(q)$[$ I(q) + L(q)$]$ +C where F(q) is the magnetic form factor of the Tb$^{3+}$ ion, L(q) is a Lorentzian function and C a constant background. The liquid-like function I(q)=$\gamma$$_1$$\cdot$ sin(qR$_1$)/(qR$_1$) accounts for correlations between first neighbor Tb pairs at a distance R$_1$. It
  has two maxima in the measured q-range, as also seen in TTO \cite{Gardner99}.
  The negative $\gamma$$_1$ value shows
  that first neighbor Tb pairs are AF coupled, and its temperature dependence (Fig. \ref{fig4.eps}) reflects the increasing correlations as temperature decreases.
  The Lorentzian term L(q) accounts for ferromagnetic correlations, expected
  when approaching the transition towards the ordered spin ice, and actually observed below \cite{Chapuis07,Mirebeau08}.
    The Lorentzian term enhances the magnetic intensity at low q's, shifts the position and damps the intensity of the two maxima. Under pressure, the $\gamma$$_1$ value
  remains unchanged, whereas the Lorentzian term vanishes (Fig. \ref{fig3.eps}b).
  The fit of the SRO under pressure is slightly improved by inserting third neighbor
  correlations between Tb moments (with ferromagnetic $\gamma$$_3$),
  noticing that second neighbor Tb pairs are absent in the pyrochlore structure \cite{Greedan91}.
  The suppression of the Lorentzian term by pressure suggests that the critical fluctuations
  associated with the ordered spin ice transition vanish, as the F ordered moment decreases and AF order is stabilized.

 The nature of the pressure induced ground state may be understood by referring both to TSO at ambient pressure and TTO under pressure and stress. At 0.06~K the ordered moment M per Tb ion, considering both F and AF structures, can be calculated as M$^2$=M$_{F}$$^2$+M$_{AF}$$^2$, yielding M= 4.3(3)\,$\mu_B$. Taking into account the long period structure, the ordered moment of which is evaluated to  1.3(2)\,$\mu_B$ by comparing the intensity of the q=1.01\AA$^{-1}$ peak to those of the (201) or (200) peaks, one gets a value of 4.5(3)\,$\mu_B$.  So the pressure induced ordered moment is strongly  reduced with respect to the ambient pressure value of 5.9(1)\,$\mu_B$. We attribute this effect to the enhancement of the spin liquid fluctuations, which should naturally wash out the magnetic order.

  We therefore conclude that the applied pressure favors both SL and AF orders at the expense of the ordered spin ice.
   To understand this result, one needs to consider the effects of isotropic and uniaxial pressure components separately. An isotropic pressure of 4.6 GPa induces an average compression of the lattice $\Delta$V$/$V of about 2\% \cite{Apetrei07}. This favors the SL rather than the ordered spin ice, as under chemical pressure when Sn is replaced by Ti of smaller ionic radius. Compressing the lattice enhances the AF superexchange with respect to the F dipolar interaction. The effect of the uniaxial component of 0.3 GPa depends on stress orientation with respect to the crystal axes, and a powder average must be performed.
    In TTO single crystal \cite{Mirebeau04}, a stress along a [110] axis induces AF order with
    ordered moment at 0.1~K up to 3.9\,$\mu_B$ and T$_{\rm N}$ of 1.8~K.
     The [110] stress relieves the geometrical frustration by inducing 3 different bond lengths in the distorted tetrahedron.  This is not the case for stresses along [111] or [100] axis, which have a much smaller effect: a [111] stress yields an ordered moment of 0.4(1) $\mu_B$ and T$_{\rm N}$  value  of 0.7(1) ~K.

    To evaluate the effect of the stress on the powder sample, noticing that the smallest angle between [110] and [111] axes is 35$^\circ$,
    we assume that only stresses closer to 
    a $[$110$]$ axis
     may induce AF order. In a powder sample, the probability
     of finding a [110] type axis in a cone of angle $\theta$ is simply given by P($\theta$) =6(1-cos$\theta$),
     for $\theta$ $\leq$30$^\circ$.
     Taking for instance $\theta$ =20$^\circ$, we find that 36\% of the grains should experience an efficient stress.
     We therefore conclude that both SL and AF orders are favored by pressure, but in different ways.
     With the above assumptions, the AF structure involves only the well-oriented grains (36\%),
      with a strong AF moment M$_{AF}$= 4.7$~\mu_B$, and a total one of 5.4~$\mu_B$, whereas the
      other grains stabilize SL and ordered spin ice states.

 Our results bring strong evidence that the magneto elastic coupling
 responsible for the pressure-induced AF order in TTO is also at play in TSO.
 The high sensitivity of the magnetic interactions to a pressure induced distortion allows one to tune ordered spin ice, spin liquid and AF orders through pressure and stress. This coupling also plays a role at ambient pressure. In TTO, a spontaneous distortion was observed, likely precursor of a Jahn-Teller transition \cite{Ruff07}. Recent results\cite{Bonville08} suggest that this distortion is also present in TSO
 and could help to stabilize the ordered spin ice state.
 As shown in ref.\onlinecite{Bonville08}, a quadratic distortion corresponding to an energy
 scale D$_Q$=~0.2~K can account for the canting angle $\alpha$=13$^\circ$, still unexplained.
 One naturally expects this distortion to increase under stress.
 This could explain both the increase of the canting angle in the F structure,
  (a distortion D$_Q$=~0.4~K yields $\alpha$=26$^\circ$ close to the experimental value),
 and the onset of the AF structure.

 The pressure induced tunability of TTO and its strong spin lattice coupling allow one to consider it as a "soft" spin ice, contrary to model spin ices. In Ho$_2$Ti$_2$O$_7$ and Dy$_2$Ti$_2$O$_7$ magnetic correlations are insensitive to pressure down to 1.4~K and up to 6~GPa\cite{Mirebeau04bis}, although their high field magnetization is slightly decreases under uniaxial pressure \cite{Mito07} and isotropic pressure may tune the charge of the magnetic monopoles\cite{Moessner08}.

  In conclusion, we have shown that applying pressure in Tb$_2$Sn$_2$O$_7$ allows one to destabilize
  the ordered spin ice state and induce spin liquid and antiferromagnetic orders.
   This onset occurs through two different mechanisms, which evidence the effect of isotropic compression on the energy balance of magnetic interactions, and the influence of  pressure induced distortion on the magnetic exchange, respectively.

    I. M.  thanks P. Bonville for comunicating unpublished results, G. Andr\'e and J. M. Mirebeau for fruitful discussions. H. Cao acknowledges support from the 
    Triangle de la Physique.

\end{document}